# Motion of 2D exciton in momentum space leads to pseudospin distribution narrowing on the Bloch Sphere

Garima Gupta[1], Kenji Watanabe[2], Takashi Taniguchi[3], and Kausik Majumdar[1,*]

[1]*Department of Electrical Communication Engineering, Indian Institute of Science, Bangalore 560012, India*

[2] *Research Center for Electronic and Optical Materials, National Institute for Materials Science, 1-1 Namiki, Tsukuba 305-0044, Japan*

[3] *Research Center for Materials Nanoarchitectonics, National Institute for Materials Science, 1-1 Namiki, Tsukuba 305-0044, Japan*

*Corresponding author, E-mail: kausikm@iisc.ac.in*

**Abstract:** Motional narrowing implies narrowing induced by motion, for example, in nuclear resonance, the thermally induced random motion of the nuclei in an inhomogeneous environment leads to counter-intuitive narrowing of the resonance line. Similarly, the excitons in monolayer semiconductors experience magnetic inhomogeneity: the electron-hole spin-exchange interaction manifests as an in-plane pseudo-magnetic field with a periodically varying orientation inside the exciton band. The excitons undergo random momentum scattering and pseudospin precession repeatedly in this inhomogeneous magnetic environment – typically resulting in fast exciton depolarization. On the contrary, we show that such magnetic inhomogeneity averages out at high scattering rate due to motional narrowing. Physically, a faster exciton scattering leads to a narrower pseudospin distribution on the Bloch sphere, implying a nontrivial improvement in exciton polarization. The in-plane nature of the pseudo-magnetic field enforces a contrasting scattering dependence between the circularly and linearly polarized excitons – providing a spectroscopic way to gauge the sample quality.

The conservation of angular momentum during light-matter interaction allows us to transfer information from the polarization vector of a photon to the pseudospin vector (**S**) of an exciton. The exciton, being composed of Coulomb-bound oppositely charged particles, is amenable for controlled manipulation through external stimuli[1–3], an advantage over the photon. For the two-dimensional (2D) excitons in the monolayers of semiconducting transition metal dichalcogenides (TMDs), information such as circular polarization (CP) and linear polarization (LP) of light can be encoded as valley polarization and valley coherence of excitons, respectively, thanks to the spin-valley locking in the monolayer[4]. Experimentally, the degree of circular and linear polarization (DOCP and DOLP) in polarization-resolved measurements provides a direct estimate of how efficiently the exciton can retain the valley information[5,6]. For practical applications, it is necessary that the exciton preserves the encoded information over a long duration. This task is highly challenging in these monolayers as the large exciton binding energy in the 2D limit due to the strong electron-hole wavefunction overlap results in a large spin exchange interaction within the exciton. The resulting in-plane pseudo-magnetic field (**Ω**) tends to degrade both valley polarization and valley coherence[7,8] at an ultrafast timescale[9,10].

This dynamics of **S** is well captured by the Maialle-Silva-Sham (MSS) equation[11]:

$$\frac{d\mathbf{S}(\mathbf{Q})}{dt} = \mathbf{\Omega}(\mathbf{Q}) \times \mathbf{S}(\mathbf{Q}) - \frac{1}{\tau}\mathbf{S}(\mathbf{Q}) + \sum_{\mathbf{Q}'} W_{\mathbf{Q}\mathbf{Q}'}\left[\mathbf{S}(\mathbf{Q}') - \mathbf{S}(\mathbf{Q})\right] + \mathbf{G} \quad (1)$$

Upon generation with a rate vector **G**, the exciton scatters from one center-of-mass (COM) momentum state (**Q**) to another (**Q**′) at a rate $W_{\mathbf{Q}\mathbf{Q}'}$ in the exciton band (Fig. 1a), leading to precession[11] of **S** around **Ω** at each **Q** (Fig. 1b-c). **Ω** represents an effective magnetic field (referred to as pseudo-magnetic field throughout this paper) arising from the exchange interaction, and is quantified as $\mathbf{\Omega} = \Omega\left[\cos(2\phi)\,\hat{x} + \sin(2\phi)\,\hat{y}\right]$ in the absence of an external magnetic field. Here $\Omega$ is the precession frequency and $\phi\left[= \tan^{-1}(Q_y/Q_x)\right]$ represents the polar angle in the momentum space[7,12]. Scattering and pseudospin precession keep repeating till the lifetime of the exciton ($\tau$), and the polarization of the photon emitted upon exciton recombination depends on the direction of **S** on the Bloch sphere at $t = \tau$.

The different terms in equation 1 readily points to the ways to improve the DOCP and DOLP. For example, a reduction in the exchange interaction (for example, through dielectric screening[7,13,14]) reduces the torque in the first term, thus suppressing the net pseudospin precession frequency. On the other hand, a reduction in $\tau$ in the second term improves polarization by reducing the total time available to the exciton for random phase accumulation[7,15] (**SI Note 1**).

However, there is a lack of understanding in the intriguing role of momentum scattering rate ($W_{QQ'}$ in equation 1) on the polarization of the 2D excitons, which is the focus of this work. We reveal two interesting observations: (1) there is a stark contrast in the DOCP and DOLP in the low scattering regime, and (2) high scattering rate counter-intuitively improves both DOCP and DOLP – providing a viable technique to maintain high valley polarization and valley coherence.

Fig. 1d shows the calculated steady-state DOCP (in blue) and DOLP (in red) as a function of momentum scattering rate (embedded in the parameter $w (\propto W_{QQ'})$, defined in **SI Note 2**). In the calculation, we have taken the elastic Coulomb scattering from charged impurities as the dominant exciton scattering mechanism at low temperature[7,9,16] (in agreement with our measurement conditions and observations, as discussed later), and hence the scattering rate is an indicator of sample quality. The plots are obtained by solving the steady-state form of equation 1 by using experimentally calibrated parameters (**SI Note 2**). We observe three striking features in Fig. 1d: (a) At small $w$ (high sample quality, left part of the yellow shaded regime), while DOLP is high, interestingly, DOCP is small. (b) With an increase in $w$ (right part of the yellow shaded regime), DOLP reduces and reaches its minimum, while DOCP improves slowly, but monotonically. (c) For large $w$ (high sample disorder, green shaded regime), both DOCP and DOLP exhibit a strong enhancement, maintaining the condition of DOLP $\geq$ DOCP in the entire scattering range.

To understand the difference between DOCP and DOLP in the low scattering regime, let us first consider $w = 0$ case. Fig. 1b-c (left column) schematically show the exciton pseudospin on the Bloch sphere for CP and LP light excitation, respectively. At time $t = 0$, the CP (LP) pseudospin is orthogonal (parallel) to the in-plane $\mathbf{\Omega}$. The CP pseudospin thus experiences maximum torque due to $\mathbf{\Omega}$, and precesses by an angle equal to $\Omega\tau$ during the exciton lifetime (see **Supporting Video 1** for the time evolution of CP pseudospin on the Bloch sphere). $\Omega\tau$ can be large in the case of monolayer TMDs due to strong electron-hole exchange interaction, forcing fast valley depolarization, and hence a low DOCP. In **SI Note 3**, we show a CP-resolved PL spectrum showing a DOCP of ~0% obtained in one of our cleanest samples. On the other hand, $\mathbf{\Omega}$ being parallel to LP pseudospin[9,17], exerts zero torque and hence the pseudospin does not precess at all (**Supporting Video 2**), maintaining 100% DOLP all along.

As $w$ becomes nonzero (middle column of Fig. 1b-c), the magnitude of the torque from $\mathbf{\Omega}$ on the CP pseudospin reduces as $\mathbf{S}$ and $\mathbf{\Omega}$ do not anymore remain orthogonal when the exciton scatters to different $\mathbf{Q}$ states. Accordingly, the DOCP improves with an increase in $w$ (**Supporting Video 3**). On the other hand, the LP exciton pseudospin experiences nonzero torque after scattering, as compared to the initial zero-torque situation. Accordingly, it accumulates random phase on precessing about $\mathbf{\Omega}$ at different $\mathbf{Q}$ states (**Supporting Video 4**), and hence the DOLP starts degrading. We denote this as the exchange-dominated regime (yellow shaded regime in Fig. 1d).

As $w$ increases further, interestingly, we observe a strong enhancement in both exciton DOCP and DOLP (green shading in Fig. 1d). This enhancement is similar to the motional narrowing phenomenon[18] observed in NMR, where the resonant linewidth narrows down due to fast nuclei motion in an inhomogeneous environment. Similarly, in our case, due to a varying $\mathbf{\Omega}$ resulting from the random exciton momentum scattering in the $\mathbf{Q}$-space, the exciton pseudospin encounters a magnetic inhomogeneity in spite of the absence of an external magnetic field. The schematic in Fig. 1b-c (right column) depicts this situation where the pseudospin hardly moves with time from its initial orientation, both for CP and LP. This counter-intuitive improvement of DOCP and DOLP occurs when the exciton momentum scattering rate is much faster than the pseudospin precession frequency (see **Supporting Video 5** for CP and **6** for LP). We denote this as the motional narrowing regime in Fig. 1d (green shading).

To further understand the role of motional narrowing, we carry out the Monte Carlo simulation within the exciton light cone (**SI Note 4**). The results are summarized in Fig. 2. We simultaneously observe the time evolution of (a) the exciton population distribution in the $\mathbf{Q}$-space, and (b) the corresponding pseudospin distribution on the Bloch Sphere, at different $w$. For LP excitation, at small $w$, as time evolves, a spread in the exciton population in the $\mathbf{Q}$-space (top panel in Fig. 2a) results in an increase in the spread of the pseudospin distribution on the Bloch sphere (bottom panel in Fig. 2a) (**Supporting Video 7**), gradually degrading DOLP. However, at high $w$, although the exciton population spreads much faster in the $\mathbf{Q}$-space (top panel in Fig. 2b), the corresponding spread in the pseudospin distribution is much slower (bottom panel in Fig. 2b) (**Supporting Video 8**), keeping DOLP intact for longer. Thus, a higher scattering rate results in a slower decay in the DOLP with time (Fig. 2c). This captures the essence of the motional narrowing effect: the pseudospin distribution on the Bloch sphere narrows down as the scattering rate (motion) increases in the momentum space. We observe similar motional narrowing for the CP excitation as well, as depicted in **SI Note 5**.

We analytically derive (**SI Note 6**) the variance of the pseudospin distribution in the motional narrowing regime as given below:

$$\sigma^2 = \tau^2 \Omega^2 \left(\frac{1}{N} + f\frac{[N-1]}{N}\right) \approx \tau^2 \Omega^2 \left(\frac{1}{N} + f\right) \quad (2)$$

Here, $N$ is the average number of scattering events in the exciton lifetime $\tau$ and is large in the motional narrowing regime. $f = \langle \cos(2\phi_i - 2\phi_j) \rangle$, where $\phi_i = \tan^{-1}(Q_{y,i}/Q_{x,i})$ is the polar angle of the exciton COM momentum after the $i$th scattering event, and the average is taken over all possible combinations ($i \neq j$). $f$ captures the magnetic inhomogeneity due to varying orientation of $\Omega$. We obtain $f$ and $N$ from the Monte Carlo simulation and plot in Fig. 2d as a function of $w$. Equation 2 readily captures the motional narrowing effect: an increase in the scattering rate results in a reduction in $f$ and $1/N$, reducing $\sigma^2$, and thus enhancing exciton DOLP in Fig. 2d.

We now experimentally explore the effect of scattering rate on DOCP and DOLP in monolayer $MoS_2$. As mentioned earlier, impurity scattering is the dominant scattering mechanism at the measurement temperature (4 K), and hence the exciton scattering strongly depends on the sample quality. The role of exciton-phonon scattering is discarded at such low temperature[7,9,16]. This is supported by noting the fact that exciton-phonon scattering is spatially homogeneous in nature, and we observe a strong spatial variation in the measured DOLP and DOCP in each of our samples (summarized in Figs. 3 and 4). The flatness of exciton DOLP variation with temperature in the low temperature regime (see **SI Note 7**) further shows that the phonon scattering does not play a dominant role at 4 K. The exciton-exciton scattering process is also negligible at the low exciton density ($\sim 10^8$ cm$^{-2}$ assuming 10% quantum efficiency) in our measurement[19,20]. Further, the low trion peak intensity observed in our samples (see **SI Note 8** for a representative emission spectrum) indicates an overall low doping, allowing us to ignore the role of exciton scattering with the free carriers (see **SI Note 9** for comparison between trion/exciton and defect/exciton intensity ratios in our samples).

We tune the exciton scattering rate through a variation in the ionized impurity concentration, which, in turn, is achieved by exploiting the sample inhomogeneity. The local density of gap states[21] varies spatially due to inhomogeneity. Also, due to its ultrathin nature, the monolayer experiences potential fluctuation from the surrounding environment, resulting in a spatial variation in the separation between conduction band and Fermi level. A higher local conduction band edge position with respect to the Fermi level exposes these charged impurity centers more[14,22] (see Fig. 3a). Excitons bound to these localized charged centers contribute to the defect peak emission in $MoS_2$ (see spectra in Fig. 3c-f and Fig. 4b-c). Thus, the defect peak intensity is an experimental analogue of the ionized impurity concentration, and hence the exciton coulomb scattering rate.

We prepare multiple samples of two different stacks: (a) hBN-$MoS_2$-hBN (stack A), and (b) $MoS_2$-hBN (stack B). The samples have varying hBN thickness, and hence represent different degrees of dielectric screening. The sample details are provided in the **Methods** section. The results are summarized in Fig. 3 for exciton DOLP, and in Fig. 4 for exciton and trion DOCP.

In Fig. 3b, we plot the experimentally obtained exciton DOLP as a function of the area under the defect peak (normalized by the area under the corresponding exciton peak), denoted by $I_D$, in the bottom axis. The results are plotted for four samples of stack A (blue shading) and two samples of stack B (green shading). For a given sample, the different points represent the data obtained from the laser spot focussed on spatially different points (Fig. 3g-k). Linear polarization resolved photoluminescence (PL) spectra (see **Methods**) for four representative data points (encircled in Fig. 3b) are shown in Fig. 3c-f, demonstrating a stark correlation between the defect peak intensity and the DOLP (see **SI Note 11** for peak fitting and logscale plots for defect intensity comparison). The overall trend has some distinct features: (1) DOLP is highly non-monotonic with $I_D$. (2) At low $I_D$ (cleaner samples), DOLP decreases with an increase in $I_D$ (Fig. 3c-d). (3) At higher $I_D$ regime (dirty samples), DOLP increases steeply with $I_D$, a signature of motional narrowing (Fig. 3d-f). (4) As a result of one-sided capping, stack B samples (green shading) with higher disorder exhibit strong motional narrowing effect, and hence consistently exhibit both higher $I_D$ and higher DOLP (blue shading).

Strikingly, we could reproduce all the experimental features from these samples in the calculated DOLP as a function of $w$ (top axis in Fig. 3b), as indicated by the red and blue solid traces covering the entire range (grey shaded region). The red (blue) trace corresponds to a higher (lower) dielectric screening and hence a lower (higher) $\Omega$, which is in excellent correlation with the thicker (thinner) hBN used in the respective samples. In Fig. 3g-k, we separately show the result from the individual samples and the model matches the trend in each sample only by changing the sample-dependent parameter $\Omega$.

The results for exciton DOCP versus $I_D$ along with the calculated trends are summarized in Fig. 4a, with two representative PL spectra shown in Fig. 4b-c. Unlike DOLP, DOCP stays small for clean samples, and monotonically increases with $I_D$ in the higher scattering rate regime – in agreement with our calculation. This suggests that one must resort to higher Coulomb scattering rate to achieve high DOCP in monolayers, thanks to the pseudospin distribution narrowing. Accordingly, due to higher disorder, stack B is again found to exhibit stronger motional narrowing effect as compared with stack A. On the other hand, similar to DOLP, a higher strength of dielectric screening helps to improve DOCP by reducing Ω (red versus blue trace).

The results for DOCP of trion from multiple stacks are summarized in Fig. 4d-e. While a detailed theoretical analysis for trion is beyond the scope of this work, the similar trend as exciton DOCP suggests that motional narrowing plays a crucial role in higher order excitonic complexes as well.

The statistical correlation between the calculation and the experimental results over multiple stacks and samples suggests a universal relationship between sample quality and the contrast between DOCP and DOLP (shaded part in Fig. 5a). We thus propose that the polarization contrast,

$$PC = \frac{\text{DOLP-DOCP}}{\text{DOLP+DOCP}} \qquad (2)$$

plotted on the right axis in Fig. 5a, is a useful metric that directly correlates with the sample quality at low temperature where impurity Coulomb scattering is dominant (also see **SI Note 12**). For example, we obtain a PC of ~87% from a clean Stack A sample (Fig. 5b) and ~11% from a stack B sample with large disorder (Fig. 5c).

Such contrasting behaviour between DOCP and DOLP, and the non-trivial improvement of polarization at higher scattering regime arising from the intricate pseudospin distribution on the Bloch sphere are expected to have interesting implications, such as, in gauging the sample quality in a non-invasive manner, and in retaining encoded quantum information through motional narrowing. Note that, in this work, we primarily discuss the exciton pseudospin dynamics at cryogenic temperature where impurity scattering is dominant. In the future, the role of other types of scattering processes on motional narrowing, such as scattering of exciton with free carriers (under gating) and with phonon (at higher temperature), and possible competition among different mechanisms will be intriguing to explore.

## Methods

### Sample preparation

We prepared the samples by obtaining MoS$_2$ bulk crystals from two different commercial vendors. The stacks are prepared first by mechanically exfoliating MoS$_2$ and hBN layers on a Polydimethylsiloxane (PDMS) film, followed by their transfer on a Si substrate covered with 285 nm thick SiO$_2$ under a microscope. All the samples went through a heating stage for improved adhesion.

### Sample characterization

The measurements are taken in a closed-cycle cryostat (Montana Instruments) at 4 K using an objective with numerical aperture of 0.5. The excitation wavelength and power used are 633 nm and 17 $\mu$W, respectively. To measure the exciton DOLP [$= (I_{H/H} - I_{V/H})/(I_{H/H} + I_{V/H})$], we tune the incoming excitation polarization using a half-wave plate and placing an analyzer in the emission path. $I_{H/H}$ ($I_{V/H}$) represents the co-(cross-) linearly polarized PL count. For the DOCP [$= (I_{\sigma+/\sigma+} - I_{\sigma-/\sigma+})/(I_{\sigma+/\sigma+} + I_{\sigma-/\sigma+})$], we place a quarter-wave plate right before the objective lens at an angle of 45° with respect to the incoming polarization direction. The instrument calibration for the polarization resolved measurements is verified using the Raman spectra of MoS$_2$ and silicon in **SI Note 10**. We obtain the DOLP and DOCP values by fitting the corresponding spectra (see **SI Note 11** for fitting of each spectrum shown in Fig. 3c-f and Fig. 4b-c).

## Supporting Information

Dependence of exciton DOLP and DOCP on the exciton lifetime, calculation details of exciton DOLP and DOCP, experimental demonstration of obtaining ~0% DOCP in hBN-capped monolayer MoS$_2$, Monte Carlo simulation details, Monte Carlo simulation results for exciton circular polarization, analytical derivation of the variance of the pseudospin distribution, temperature dependent exciton DOLP in monolayer MoS$_2$ on hBN, representative PL spectrum of hBN-capped monolayer MoS$_2$ showing weak trion peak intensity, Trion/exciton versus defect/exciton ratio, system calibration data for polarization resolved measurement, curve fittings for the PL spectra shown in this manuscript, polarization contrast as a function of momentum scattering rate at varying exciton lifetimes.

**Acknowledgements**

K.W. and T.T. acknowledge support from the JSPS KAKENHI (Grant Numbers 21H05233 and 23H02052) and World Premier International Research Center Initiative (WPI), MEXT, Japan. K.M. acknowledges the support from a grant from Science and Engineering Research Board (SERB) under Core Research Grant, grants from the Indian Space Research Organization (ISRO), a grant under SERB TETRA, and a seed funding from Quantum Research Park (QuRP), funded by Karnataka Innovation and Technology Society (KITS), K-Tech, Government of Karnataka.

**Competing Interests**

The authors declare no competing interests.

**Data Availability**

Data available from the corresponding author upon reasonable request.

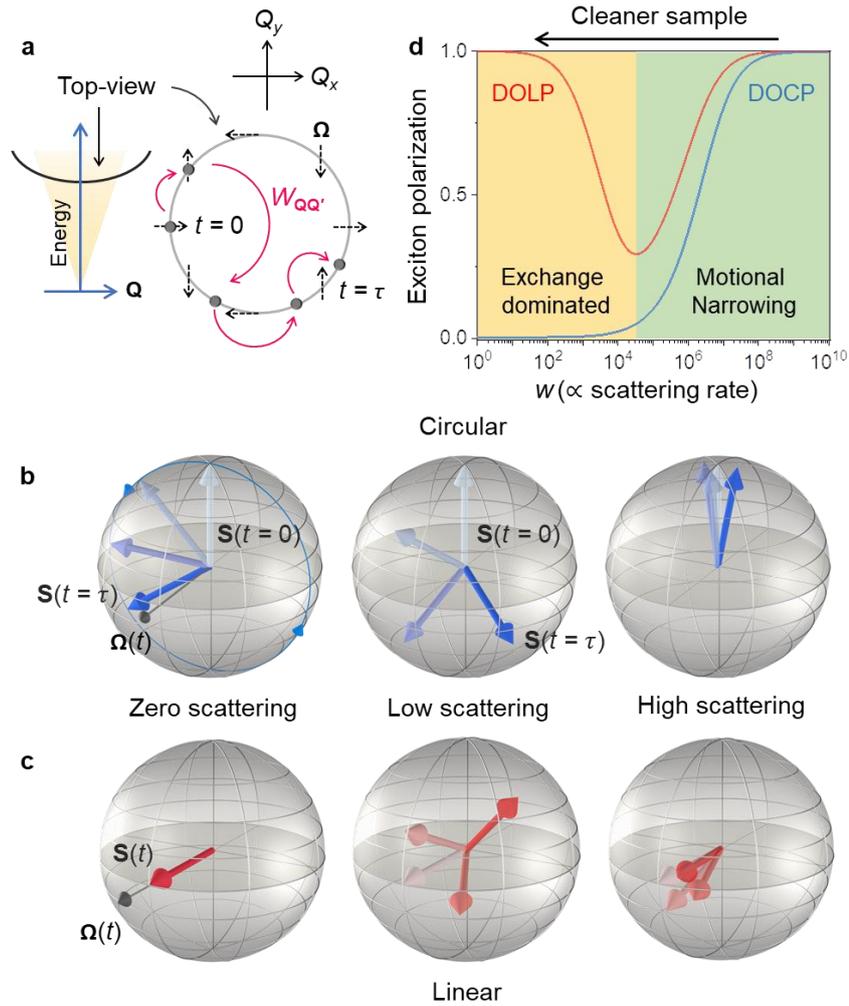

**Figure 1: Evolution of exciton polarization with scattering rate – DOCP versus DOLP.** (a) Left panel: Exciton band dispersion as a function of **Q**. The yellow shaded region represents the light cone. Right panel: The top view of a ring within the light cone. This schematic demonstrates an exciton undergoing elastic Coulomb scattering ($W_{QQ'}$, in pink arrows) in the exciton band from generation ($t = 0$) till its lifetime ($t = \tau$). The orientation of the exchange field $\mathbf{\Omega}$ varies with a winding number of two around the circle. (b-c) Schematic Bloch sphere representation of the time evolution of exciton pseudospin for (b) circularly polarized and (c) linearly polarized light excitation at zero scattering rate (first column), low scattering rate [middle column, corresponding to yellow shading in (d)], and at high scattering rate [right column, corresponding to green shading in (d)]. The gradient in the colour of the pseudospin Bloch vector (from light to dark) represents progressing time. (d) Calculated trend in exciton DOLP (in red) and DOCP (in blue) with momentum scattering rate in the monolayer. The yellow and green shading distinguishes the low scattering (exchange dominated) and high scattering (motional narrowing) regimes.

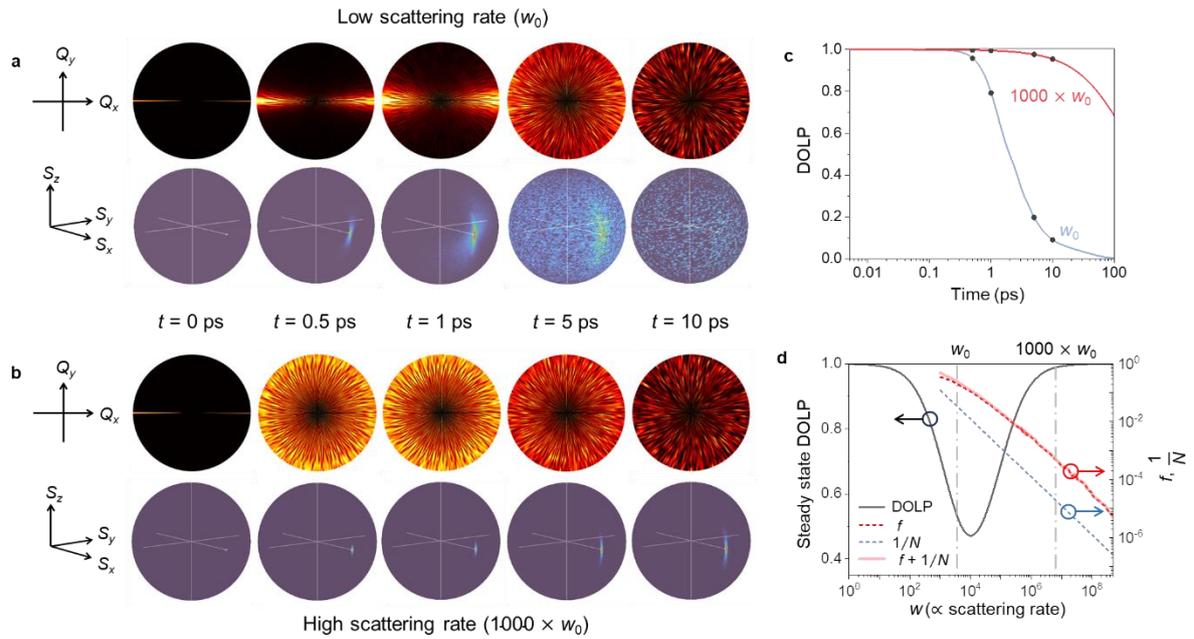

**Figure 2: Monte Carlo simulation showing pseudospin distribution narrowing on the Bloch sphere due to motion in the momentum space.** (a-b) Monte Carlo simulation at (a) low and (b) high momentum scattering rate. The top panels in both (a) and (b) show the time evolution of the exciton population distribution in the center-of-mass momentum space. The bottom panels represent the corresponding exciton pseudospin distribution on the Bloch sphere. The axes are shown in the left inset of (a) and (b). (c) The calculated exciton DOLP as a function of time at the two scattering rates used in (a-b). The black circles are the time points corresponding to the Monte Carlo plots in (a-b). (d) Calculated steady state exciton DOLP (black line) as a function of momentum scattering rate. Line plots showing the trend of $f$ (red dashed line), $1/N$ (blue dashed line), and $f + 1/N$ (red solid line) as a function of momentum scattering rate. The vertical dashed-dotted lines indicate the scattering rates chosen in (a-c).

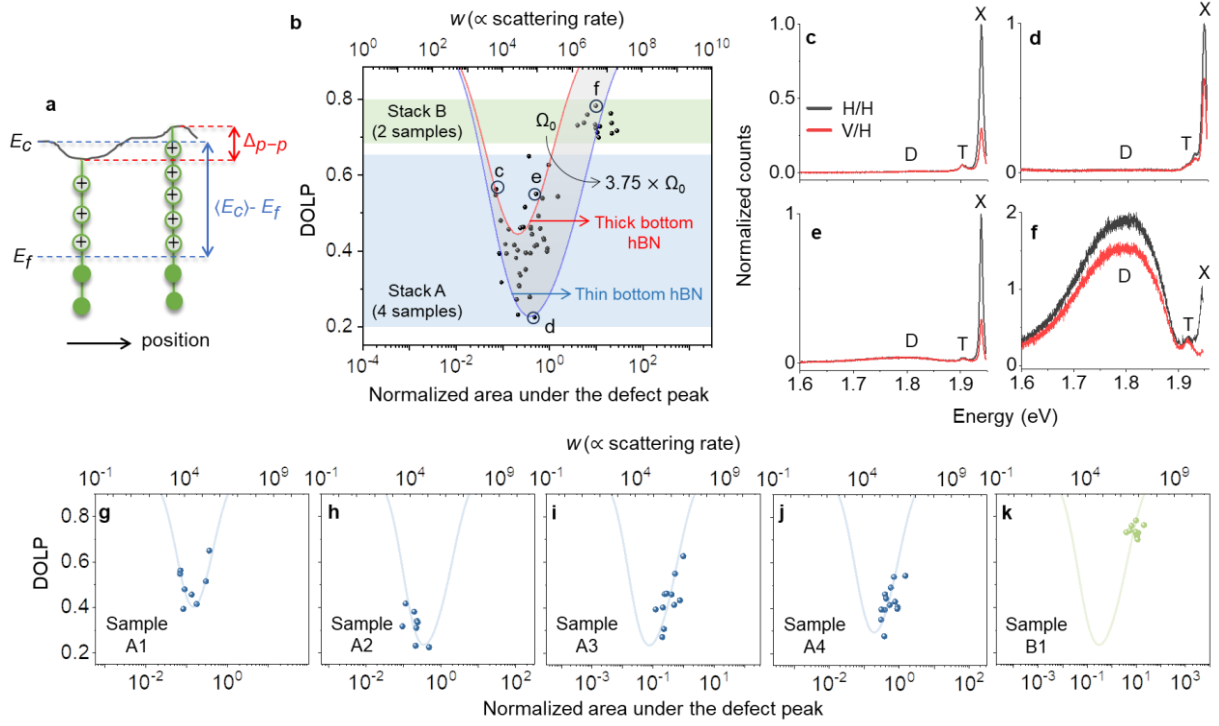

**Figure 3: Universal trend of exciton valley coherence.** (a) Schematic representation of varying concentration of ionized impurity atoms in the 2D real space. The uncompensated ionized defects lie above the Fermi level (in open circles with plus sign), whereas the compensated charge neutral defects lie below the Fermi level (in solid circles). Due to inhomogeneity induced local fluctuations in the separation between the conduction band minimum ($E_c$), and the Fermi level ($E_f$), the concentration of the uncompensated ionized defects varies from one point to another. Such potential fluctuations (peak to peak variation denoted by $\Delta_{p-p}$) are however much smaller compared to the mean separation $\langle E_c \rangle - E_f$, due to low electron doping in our samples. (b) Black symbols represent the measured DOLP at different spots from four different samples of Stack A (hBN-MoS$_2$-hBN, points lying in the blue shaded region) and two different samples of Stack B (MoS$_2$-hBN, points lying in the green shaded region). The bottom axis represents the normalized defect intensity, which is the ratio of the area under the defect peak and the corresponding exciton peak. The overall experimental trend is fitted by overlapping the calculated DOLP as a function of scattering rate (top axis) by varying the exchange parameter [$\Omega = \Omega_0$ (higher screening, red trace) to $\Omega = 3.75 \times \Omega_0$ (reduced screening, blue trace)] in our simulations. (c-f) Representative linear polarization resolved photoluminescence spectra (co-polarized in black and cross-polarized in red) for the data points circled in (b). X, T, and D represent the exciton, trion, and defect peak, respectively. (g-k) Measured DOLP (symbols) from different points of individual samples (sample number

in the insets) and corresponding fitting with the model (solid trace), obtained by varying only the exchange parameter (shown in the insets).

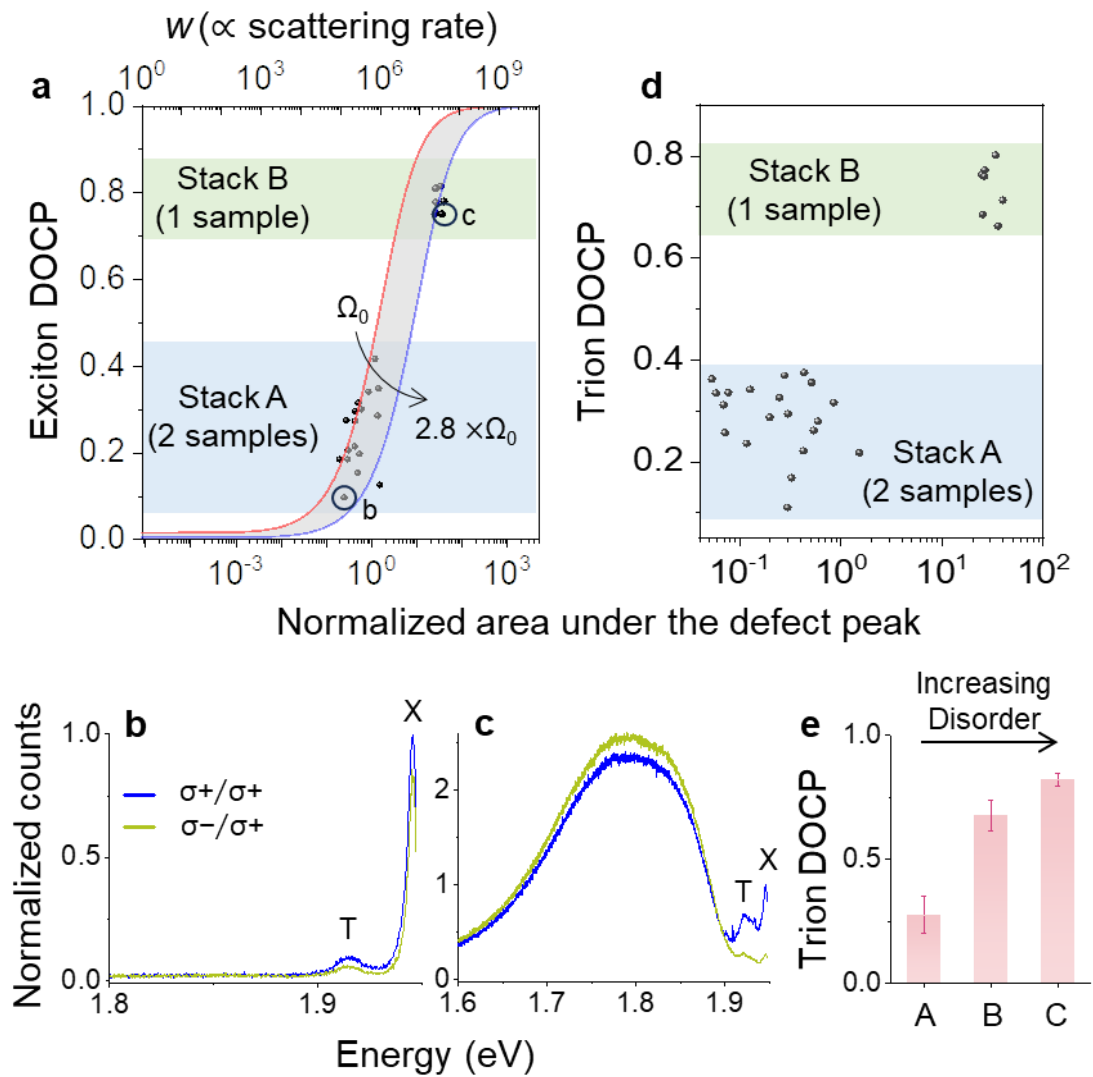

**Figure 4: Universal trend in exciton and trion valley polarization.** (a-b) Symbols represent the measured exciton DOCP at different spots from Stack A (hBN-MoS$_2$-hBN, blue shaded region, two samples) and Stack B (MoS$_2$-hBN, green shaded region, one sample). The solid red (higher screening) and blue (lower screening) traces are the fitted DOCP as a function of scattering rate (top axis). The exchange parameters used for the red and blue traces are shown in the plot. (b-c) Representative circular polarization resolved PL spectra corresponding to the points circled in (a). The blue and green traces indicate co- and cross-polarized spectra, respectively. (d) Symbols represent the measured trion DOCP at different spots from Stack A and Stack B. (e) Bar graph with an error bar comparing the average trion DOCP in stack A, B, and C (MoS$_2$-SiO$_2$), indicating an increasing trion DOCP with disorder.

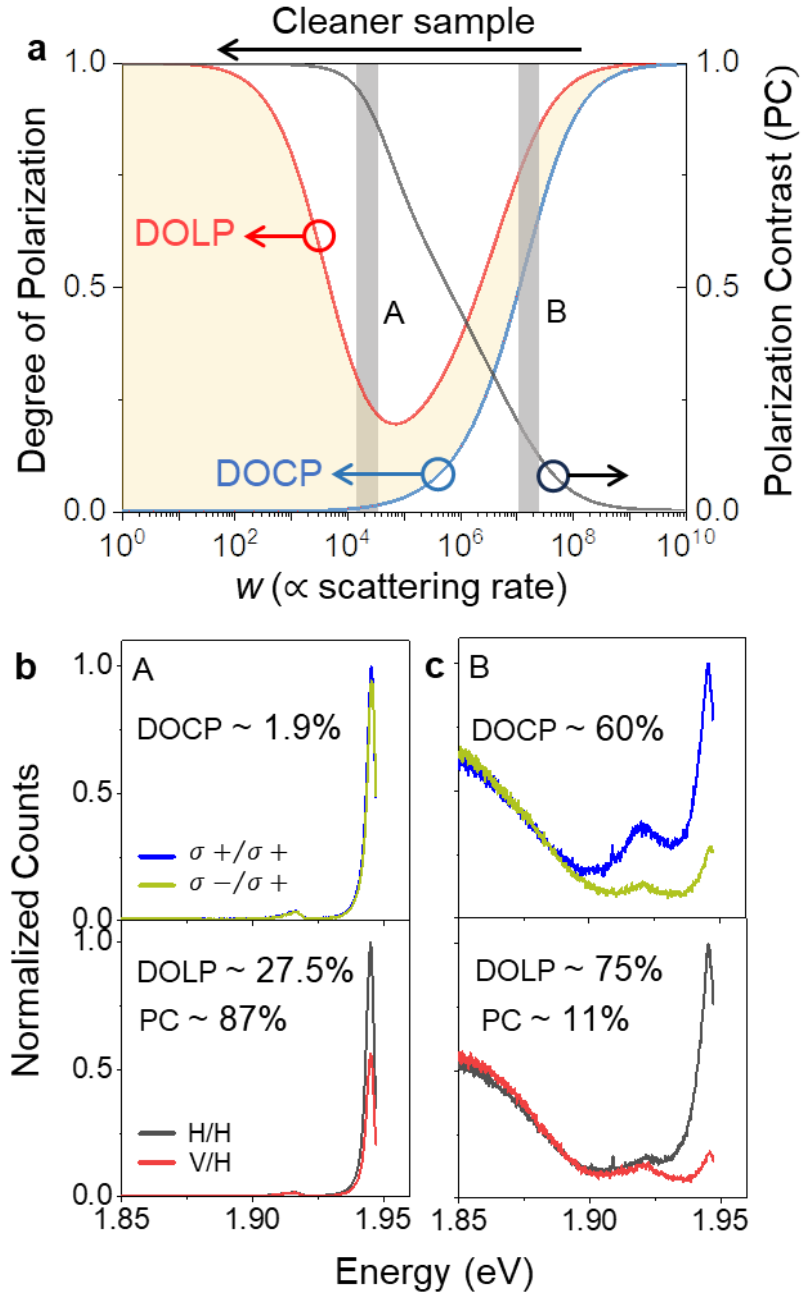

**Figure 5: Polarization contrast as a metric for sample quality.** (a) Calculated exciton DOLP (in red) and DOCP (in blue) on the left axis, and the contrast between DOLP and DOCP (in black) on the right axis, plotted as a function of the Coulomb scattering rate. The polarization contrast (yellow shading) increases monotonically with sample cleanliness (reduced scattering rate). (b-c) Circular (top row) and linear (bottom row) polarization resolved PL spectra for two specific samples, (b) one from Stack A and (c) the other from Stack B. The highlighted grey regions in (a) show the regime where these two samples lie. We obtain a higher PC of ∼ 87% from (b) the relatively cleaner stack A sample as compared to (c) the stack B sample, that gives a PC of only ∼ 11%.

TOC Graphic:

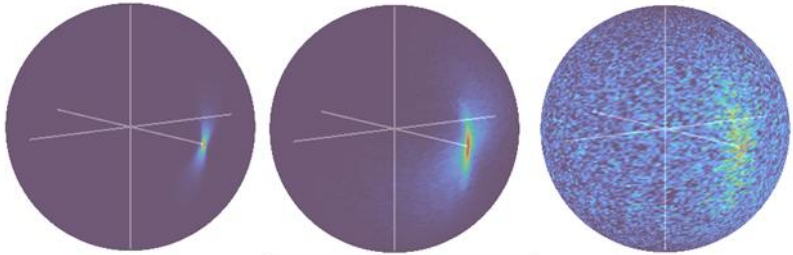
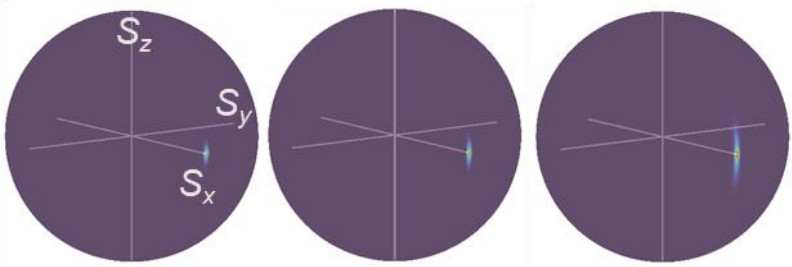

# Supporting Information:
# Motion of 2D exciton in momentum space leads to pseudospin distribution narrowing on the Bloch Sphere

Garima Gupta[1], Kenji Watanabe[2], Takashi Taniguchi[3], and Kausik Majumdar[1,*]

[1]*Department of Electrical Communication Engineering, Indian Institute of Science, Bangalore 560012, India*

[2] *Research Center for Electronic and Optical Materials, National Institute for Materials Science, 1-1 Namiki, Tsukuba 305-0044, Japan*

[3] *Research Center for Materials Nanoarchitectonics, National Institute for Materials Science, 1-1 Namiki, Tsukuba 305-0044, Japan*

*Corresponding author, E-mail: kausikm@iisc.ac.in*

**Note 1.**

**Dependence of exciton DOLP and DOCP on the exciton lifetime**

The steady-state form of the time-dependent MSS equation (Eqn. 1 in the main manuscript) is given by:

$$\mathbf{G} = \frac{1}{\tau}\mathbf{S}(\mathbf{Q}) - \mathbf{\Omega}(\mathbf{Q}) \times \mathbf{S}(\mathbf{Q}) - \sum_{\mathbf{Q'}} \underbrace{\frac{w}{Q^2 \sin^2 \frac{\alpha}{2}}}_{W_{\mathbf{QQ'}}} [\mathbf{S}(\mathbf{Q'}) - \mathbf{S}(\mathbf{Q})] \qquad (1)$$

The terms in the above equation are defined in the main manuscript. The momentum scattering rate expression $W_{\mathbf{QQ'}} = w/Q^2 \sin^2 \frac{\alpha}{2}$ ($\alpha$ represents the angle between the initial ($\mathbf{Q}$) and final ($\mathbf{Q'}$) state during scattering) corresponds to scattering of excitons with the charged impurities (see the supplementary information of ref. [1] for derivation). Using the above equation, we calculate the steady-state exciton DOLP (Fig S1.1) and DOCP (Fig. S1.2) as a function of momentum scattering rate scaling factor $w$, for different values of exciton lifetime $\tau$. The exchange frequency remains constant in the calculation.

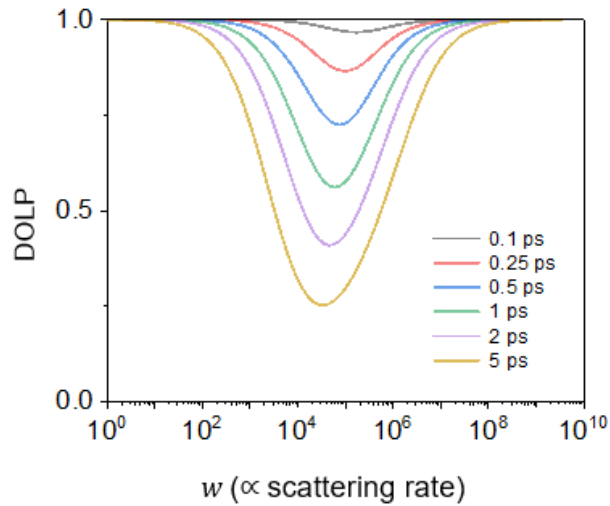

**Fig. S1.1:** Evolution of exciton DOLP as a function of momentum scattering rate at different exciton lifetimes (shown in the inset).

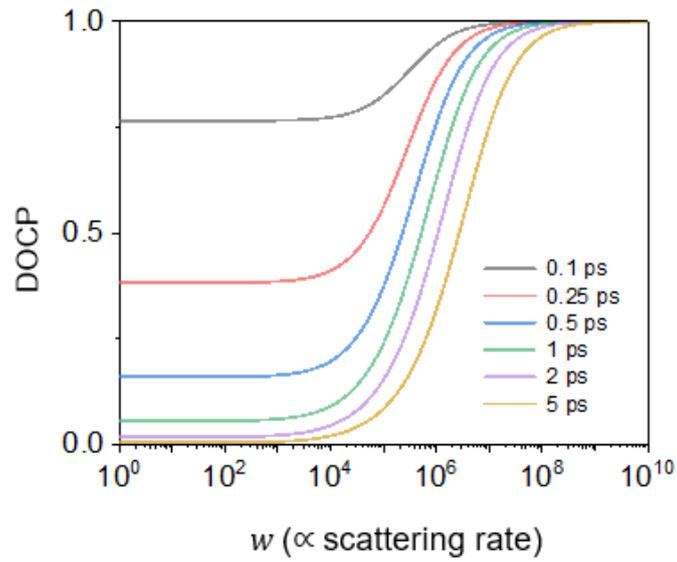

**Fig. S1.2:** Evolution of exciton DOCP as a function of momentum scattering rate at different exciton lifetimes (shown in the inset).

Thus, exciton valley polarization and valley coherence improve as the exciton lifetime reduces.

**Note 2.**

**Calculation details of exciton DOLP and DOCP**

The steady-state exciton DOLP ($\langle S_x \rangle$) or DOCP ($\langle S_z \rangle$) is obtained by solving for **S** using the following steady-state MSS vector equation[1]:

$$\mathbf{G} = -\mathbf{\Omega}(\mathbf{Q}) \times \mathbf{S}(\mathbf{Q}) + \frac{1}{\tau}\mathbf{S}(\mathbf{Q}) - \sum_{\mathbf{Q}'} \underbrace{\frac{w}{Q^2 \sin^2\frac{\alpha}{2}}}_{W_{\mathbf{QQ}'}} [\mathbf{S}(\mathbf{Q}') - \mathbf{S}(\mathbf{Q})]$$

**G** is the generate rate vector which is taken as either [1,0,0] or [0,0,1] depending on whether the excitation light is linearly or circularly polarized, respectively. The calculation of $\mathbf{\Omega}$ for the lowest energy $1s$ exciton, and the derivation of the expression for exciton-charged impurity momentum scattering rate $W_{\mathbf{QQ}'} = w/Q^2\sin^2\frac{\alpha}{2}$ is taken from ref. [1]. $\alpha$ is the angle between the initial (**Q**) and the final (**Q**′) state during scattering. We vary the scaling factor $w$ to obtain a varying exciton momentum scattering rate. The exciton lifetime has been taken to be $\tau = 3$ ps in our calculations[2–4].

**Note 3.**

**Demonstrating DOCP ~ 0% in hBN-capped monolayer MoS$_2$**

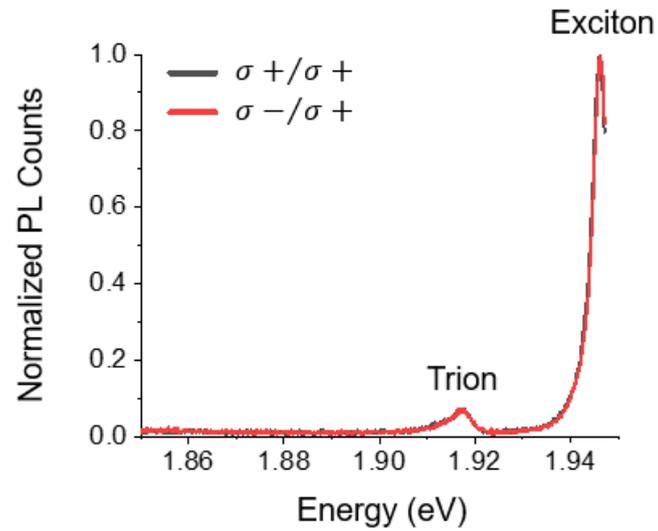

**Fig. S3.1:** Circular polarization resolved PL spectrum of a clean hBN-capped monolayer MoS$_2$ showing ~0% DOCP at 4 K. The fitted FWHM of the exciton peak is 4 meV. The excitation laser used is at 633 nm. The long-pass filter cuts off the higher energy part of the spectrum due to near-resonant excitation.

Note 4.

**Monte Carlo Simulation Details**

For the Monte Carlo simulation (Fig. 2 in the main manuscript), we generate $10^4$ excitons at each $\mathbf{Q}$ state. We only generate excitons at $\mathbf{Q}$ states within the exciton light cone. For linearly polarized excitation, we generate excitons at $\phi = 0^0$ and $180^0$ only, where $\phi$ is the polar angle in the exciton band in the momentum space. For circularly polarized excitation, the excitons are generated at all $\phi$ values. The steps in the Monte Carlo calculation of obtaining simultaneous exciton distribution in the momentum space and the Bloch sphere, as a function of time are as follows:

- **Calculation of free flight time and the exciton pseudospin precession between two scattering events**

An exciton at any state $\mathbf{Q}$ stays at the same state and undergoes scattering after a free flight time given by

$$t_f = -\frac{\ln r_1}{W(\mathbf{Q})}$$

$r_1$ represents a random number with a uniform distribution between 0 and 1. $W(\mathbf{Q})$ is the total momentum scattering rate at $\mathbf{Q}$ given by $W(\mathbf{Q}) = \sum_{\mathbf{Q}'} W_{\mathbf{QQ}'}$. During this time of flight, the exciton pseudospin undergoes presession about $\mathbf{\Omega}(\mathbf{Q})$. We obtain the modified $S_x, S_y$ and $S_z$ during the free flight time by using the cross product $\mathbf{\Omega}(\mathbf{Q}) \times \mathbf{S}(\mathbf{Q})$.

- **Selection of the new momentum state after scattering**

The final state after scattering is obtained by:

$$\Lambda_{\mathbf{Q},\mathbf{Q}'} < r_2 \leq \Lambda_{\mathbf{Q},\mathbf{Q}'+d\mathbf{Q}'}$$

$r_2$ is a another uniformly distributed random number between 0 and 1. $\Lambda_{\mathbf{Q}'}$ is defined as the successive summations from the first $\mathbf{Q}_1$ state till $\mathbf{Q}'$ normalized by the total scattering rate at $\mathbf{Q}$.

$$\Lambda_{\mathbf{Q},\mathbf{Q}'} = \frac{\sum_{\mathbf{Q}_i=\mathbf{Q}_1}^{\mathbf{Q}'} W_{\mathbf{QQ}_i}}{W(\mathbf{Q})}$$

- **Determination of the exciton lifetime**

The above two steps of free flight (and thus precession), followed by scattering are repeated till the lifetime of each exciton. The lifetime of the exciton is obtained by

$$t = -\tau \ln r_3$$

where $r_3$ is another uniformly distributed random number between 0 and 1. Here $\tau$ is the statistical exciton lifetime. We take $\tau = 3$ ps, typical exciton lifetime in monolayer TMDs.

**Note 5.**

**Monte Carlo simulation results for exciton circular polarization**

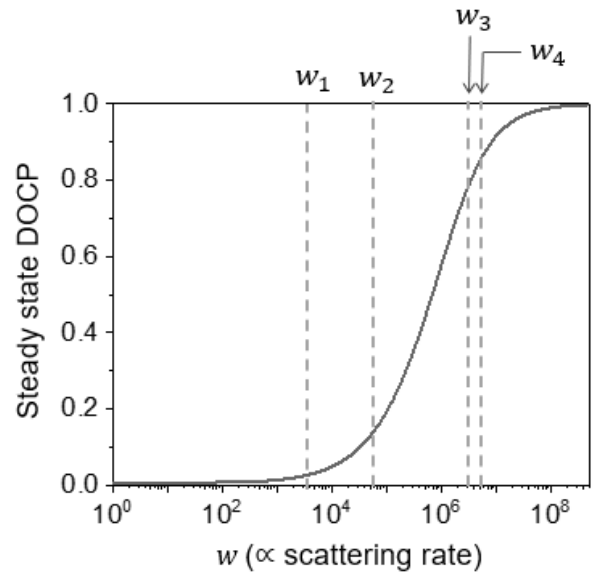

**Fig. S5.1:** Evolution of steady-state exciton DOCP as a function of momentum scattering rate. The four different values of momentum scattering rates ($w_1, w_2, w_3, w_4$) are chosen for further analysis.

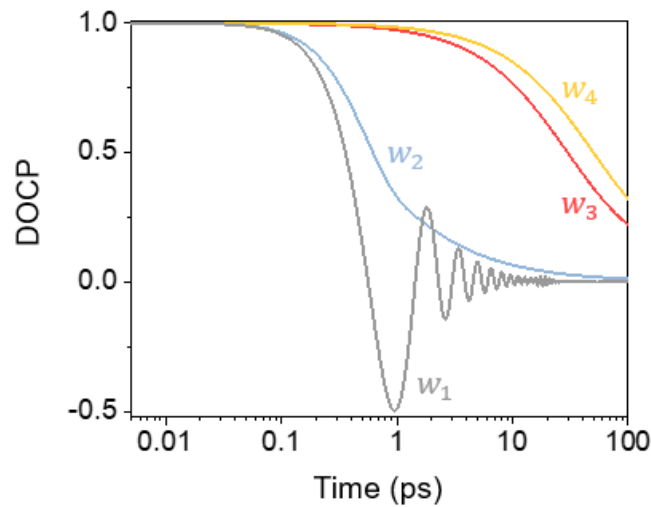

**Fig. S5.2:** Time-dependent evolution of exciton DOCP at the four different scattering rates highlighted in Fig. S5.1. Note that the oscillation at low scattering rates takes place due to pseudospin precession around $\mathbf{\Omega}$ (see **Supporting video 1**). This happens when electron-hole spin exchange effect dominates over exciton momentum scattering.

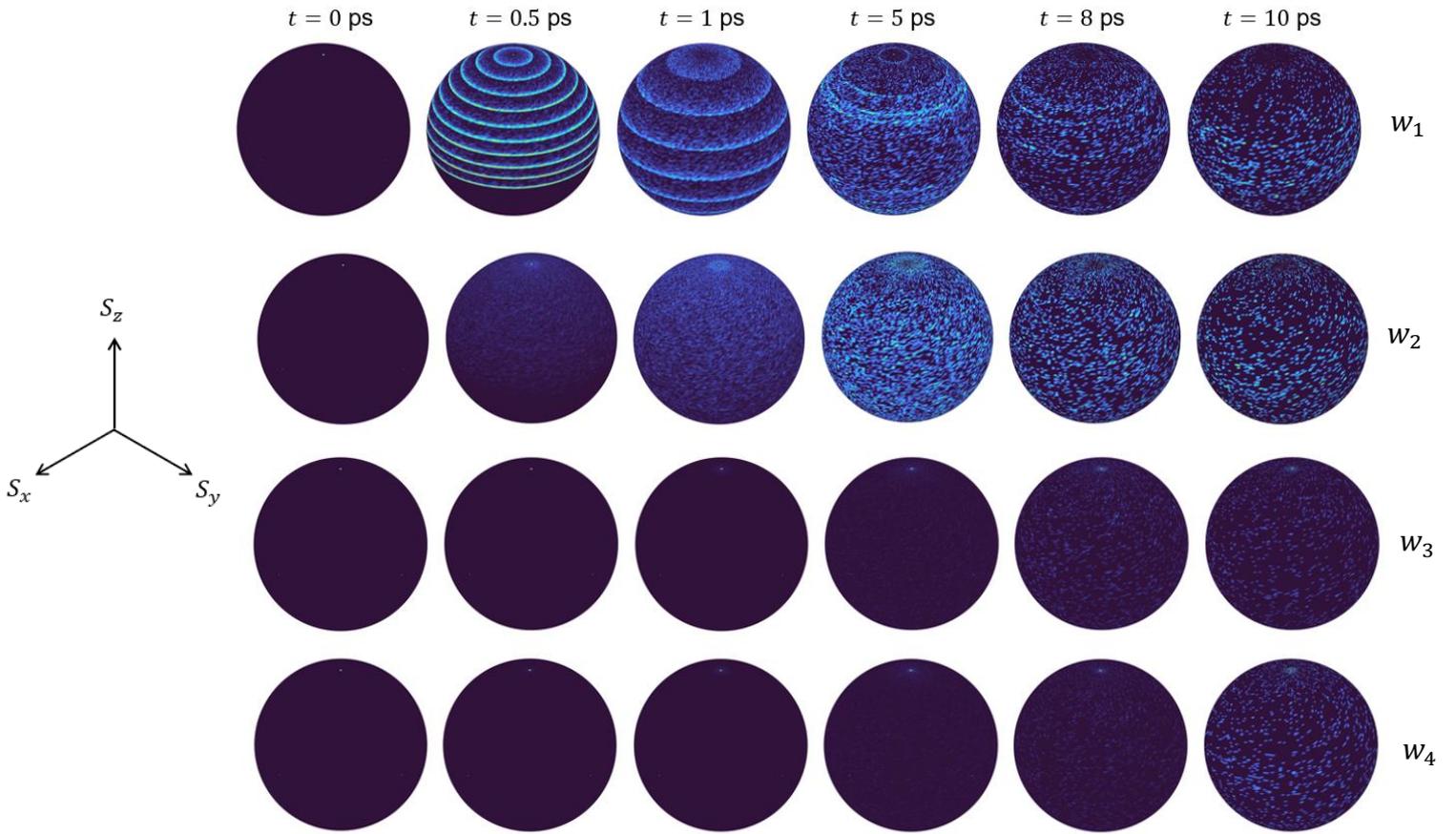

**Fig. S5.3:** Time evolution of the pseudospin phase distribution of the $10^4$ circularly polarized generated excitons on the Bloch sphere obtained using the Monte Carlo simulation. The Bloch sphere axis orientation is shown on the left. The momentum scattering rate increases from the top row ($w_1$) to the bottom row ($w_4$). Note that the rings observed on the Bloch sphere at low scattering rate ($w_1$) appear due to the $Q$ discretization in the calculation. Since $\Omega$ increases with $Q$, the different rings (corresponding to excitons residing at different $Q$ states and precessing at different exchange frequency) appear. On the other hand, at large scattering rate (for example, $w_3$ and $w_4$), the narrowing of the pseudospin distribution around the north pole is evident, indicating motional narrowing effect for circular polarization.

**Note 6.**

**Analytical derivation of the variance of the pseudospin distribution on the Bloch sphere in the motional narrowing regime**

In between the $(i-1)$ and the $i^{th}$ scattering event, an exciton at a state $\phi_i$ [$= \tan^{-1}(Q_{y,i}/Q_{x,i})$] in the **Q**-space, precesses about $\mathbf{\Omega}$ oriented along $\hat{n}_i$ [$= \cos(2\phi_i)\,\hat{x} + \sin(2\phi_i)\,\hat{y}$]. Here $\Omega$ is the precession frequency. Before the next scattering event, the pseudospin rotates by an angle

$$\vec{\varphi}_i = \Omega \tau_i \hat{n}_i$$

on undergoing precession for scattering time $\tau_i$. After many such scatterings, the pseudospin collects the rotation around various $\hat{n}_i$. Usually, rotations cannot be added vectorially as it does not commute with each other. However, this condition relaxes when the magnitude of rotation is infinitesimally small. In the scenario of commuting infinitesimal rotations, a vector addition of the rotations is possible.

In our scenario, the rotations are infinitesimal when $\Omega \tau_i \ll 1$, a condition that holds in the motional narrowing regime. Hence, the net rotation in the exciton pseudospin after $N$ scattering events in the exciton lifetime is given by

$$\vec{\varphi} = \sum_{i=1}^{N} \vec{\varphi}_i = \sum_{i=1}^{N} \Omega \tau_i \hat{n}_i$$

For multiple such excitons, the variance in the magnitude $\varphi \, (= |\vec{\varphi}|)$ provides us the information about the extent of the pseudospin phase distribution on the Bloch Sphere, given as follows:

$$\sigma^2 = \langle \varphi^2 \rangle - \langle \varphi \rangle^2$$

As $\langle \varphi \rangle^2$ goes to 0, the above equation becomes:

$$\sigma^2 = \left\langle \left( \sum_{i=1}^{N} \vec{\varphi}_i \right)^2 \right\rangle$$

$$\sigma^2 = \sum_{i=1}^{N} \langle \varphi_i^2 \rangle + 2 \left\langle \sum_{i=1}^{N} \sum_{\substack{j=1 \\ j \neq i}}^{N} \vec{\varphi}_i \cdot \vec{\varphi}_j \right\rangle$$

$$\sigma^2 = \Omega^2 \sum_{i=1}^{N} \langle \tau_i^2 \rangle + 2\Omega^2 \langle \sum_{i=1}^{N} \sum_{j=1, j \neq i}^{N} \tau_i \tau_j \, \hat{n}_i . \hat{n}_j \rangle$$

Since the lifetime $\tau_i$ and rotation axis $\hat{n}_i$ are independent of each other, the above equation is simplified as:

$$\sigma^2 = N\Omega^2 \langle \tau_i^2 \rangle + \Omega^2 N(N-1) \langle \tau_i \tau_j \rangle_{i \neq j} \langle \cos(2\phi_i - 2\phi_j) \rangle_{i \neq j}$$

We define $\langle \tau_i^2 \rangle = \tau_{rms}^2$, $\langle \tau_i \tau_j \rangle_{i \neq j} = \bar{\tau}^2$ (where $\bar{\tau} = \langle \tau_i \rangle$, since $\tau_i$ and $\tau_j$ are independent of each other), and $\langle \cos(2\phi_i - 2\phi_j) \rangle_{i \neq j} = f$.

$$\sigma^2(\varphi^2) = N\Omega^2 \tau_{rms}^2 + \Omega^2 N(N-1) \bar{\tau}^2 f$$

Approximating $\tau_{rms} \approx \bar{\tau} \approx \tau_s$ in the limit of very high scattering rate, we get

$$\sigma^2 \approx N^2 \Omega^2 \tau_s^2 \left( \frac{1}{N} + f \frac{[N-1]}{N} \right)$$

The term $N\tau_s$ can be approximated to be equal to exciton lifetime $\tau$. Hence,

$$\sigma^2(\varphi^2) \approx \tau^2 \Omega^2 \left( \frac{1}{N} + f \frac{[N-1]}{N} \right)$$

**Note 7.**

**Temperature dependence of exciton DOLP in monolayer MoS$_2$ on hBN**

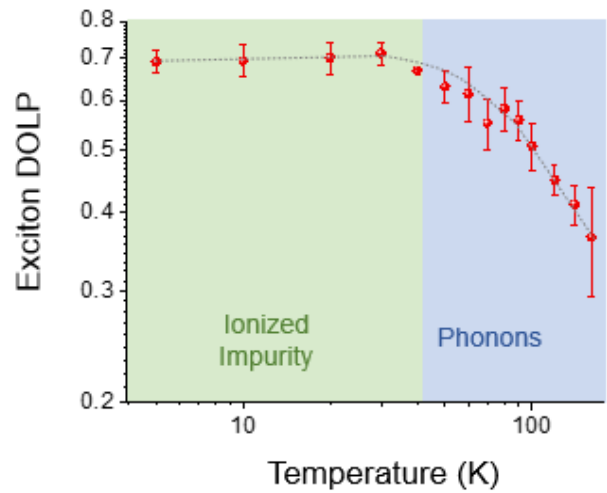

**Fig. S7.1:** Temperature dependence of exciton DOLP in monolayer MoS$_2$ on hBN. The green and the blue regions highlight the impurity scattering and phonon scattering dominated regions. The dashed black line is drawn as a guide to the eye.

**Note 8.**

**Representative photoluminescence spectrum of hBN-capped monolayer MoS$_2$ showing weak trion peak intensity taken with 633 nm excitation**

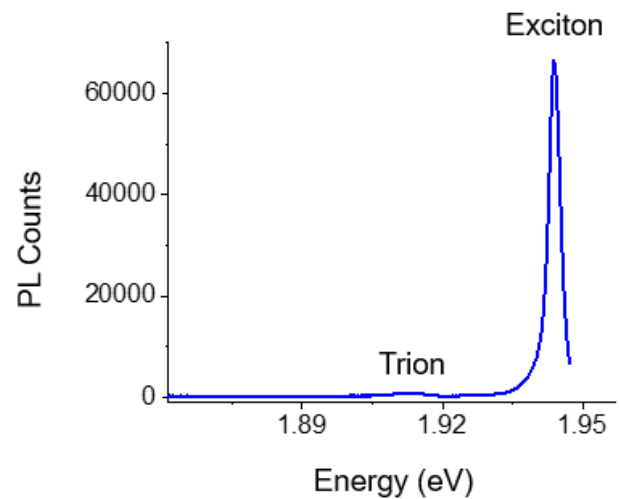

**Fig. S8.1:** A representative PL spectrum of monolayer MoS$_2$ capped with hBN on top and bottom showing weak trion peak as compared to the exciton peak at 4K (taken with 633 nm excitation laser). This implies that the exciton scattering with free carrier can be neglected in our samples at low temperature. The fitted FWHM of the exciton peak in this spectrum is 2.9 meV.

**Note 9.**

**Trion/exciton versus defect/exciton ratio**

Table S9.1 below compares the trion and defect intensity (normalized with the exciton intensity) in different stacks (stack A: monolayer $MoS_2$ sandwiched between two hBN flakes; stack B: monolayer $MoS_2$ on hBN)

|  | Trion/Exciton | Defect/Exciton | Defect/Trion |
|---|---|---|---|
| Stack A (Average taken over 39 points) | 0.069 | 0.375 | 5.44 |
| Stack B (Average taken over 10 points) | 0.106 | 9.987 | 94.039 |
| Stack B/Stack A | 1.54 | 26.6 |  |

**Table S9.1:** Comparison of average trion and defect peak intensity (both normalized with the exciton peak intensity) between stack A and stack B.

We can infer the following from the table above:

(a) The trion to exciton ratio in both the samples in much smaller as compared to the defect to exciton ratio. This implies that it is more probable for an exciton to scatter with an ionized impurity rather than a free electron (trion intensity directly reflecting electron density).

(b) When we go from stack A to stack B, the enhancement in the trion intensity (~1.54 times) is negligible as compared to the defect intensity (~26.6 times). This validates the main observation in this paper - a strong enhancement in the exciton DOLP (and DOCP) from stack A to stack B, due to motional narrowing induced by strong impurity scattering, which cannot be explained by free electron scattering as electron density remains similar in both the stacks.

**Note 10.**

**System calibration data for polarization resolved measurement**

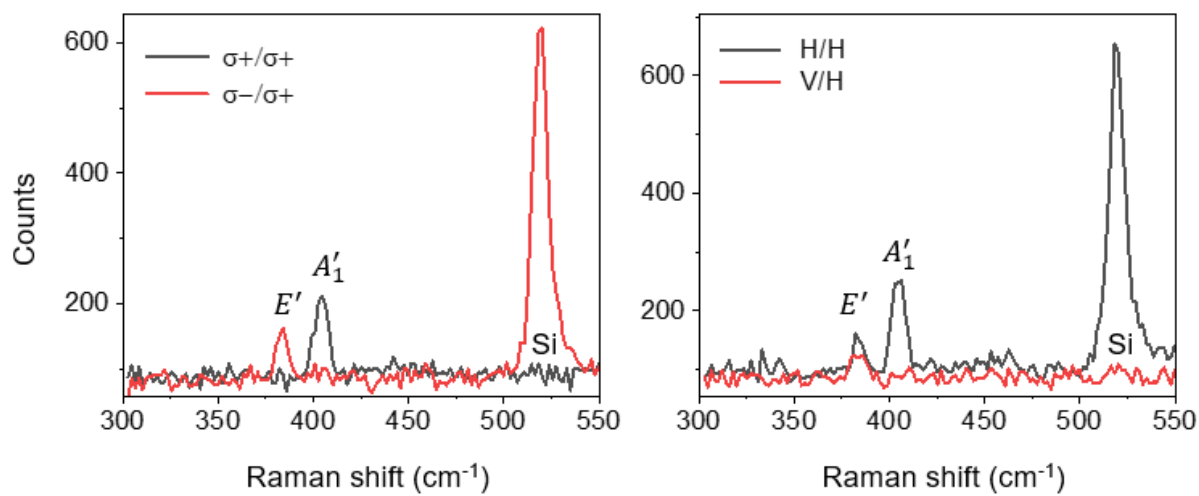

**Fig. S10.1: Left panel:** Circular polarization resolved Raman spectrum of monolayer $MoS_2$ showing an almost fully co-polarized $MoS_2$ $A'_1$ Raman peak and a cross-polarized $MoS_2$ $E'$ peak. The Si Raman peak is also fully cross-polarized[5]. **Right panel:** Linear polarization resolved Raman spectrum of monolayer $MoS_2$ showing an almost fully co-polarized $MoS_2$ $A'_1$ Raman peak. Note that, $MoS_2$ $E'$ Raman mode is not expected to be co-linearly polarized due to its Raman tensor[6]. The Si Raman peak is fully co-polarized. All the Raman spectra have been acquired at 295 K.

**Note 11.**

The fittings for Figs. 3c-f and 4b-c of main manuscript. For each spectrum, the top row shows the polarization resolved PL plots [in linear (left panel) and log (right panel) scale]. The bottom row shows the fitting of the corresponding co-polarized (left panel) and cross-polarized (right panel) spectrum in linear scale. The fitting is shown in a zoomed-in scale for better clarity.

Spectrum in Main manuscript Fig. 3c:

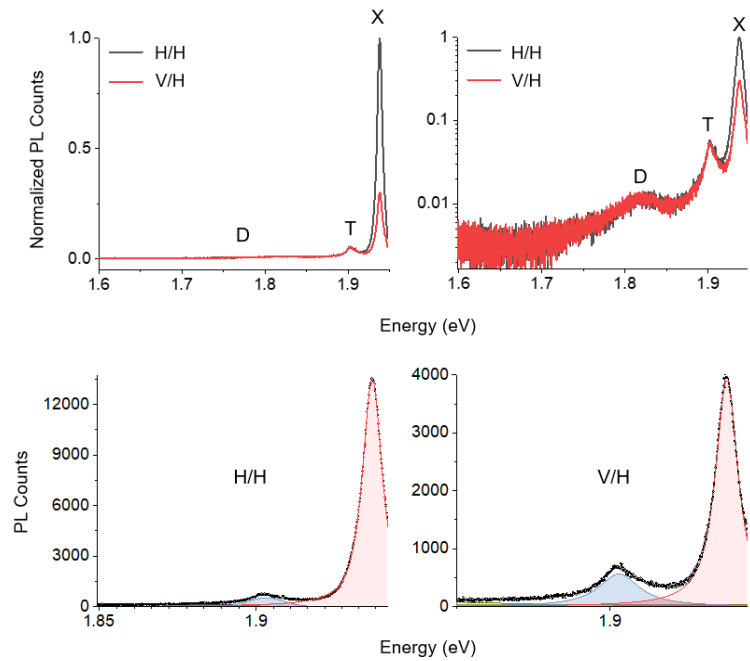

Spectrum in main manuscript Fig. 3d:

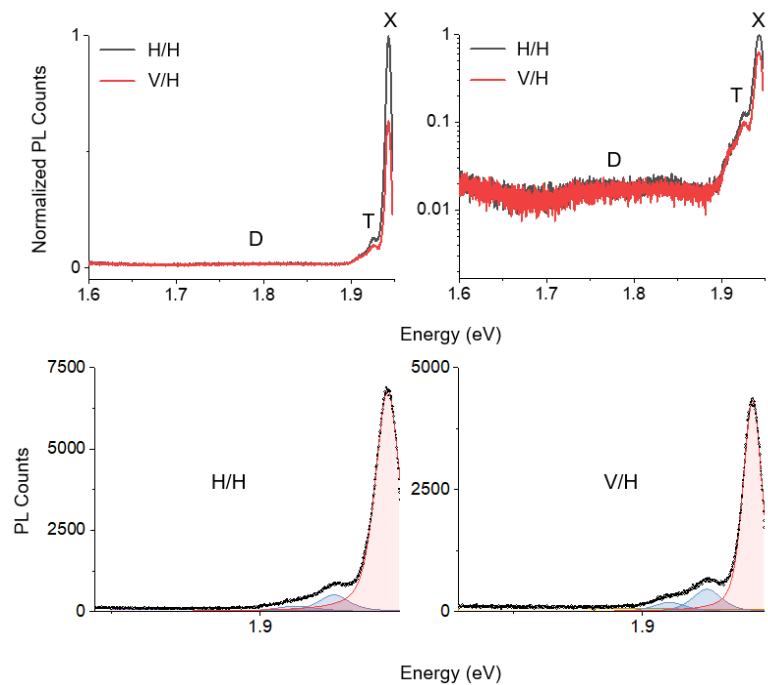

Spectrum in main manuscript Fig. 3e:

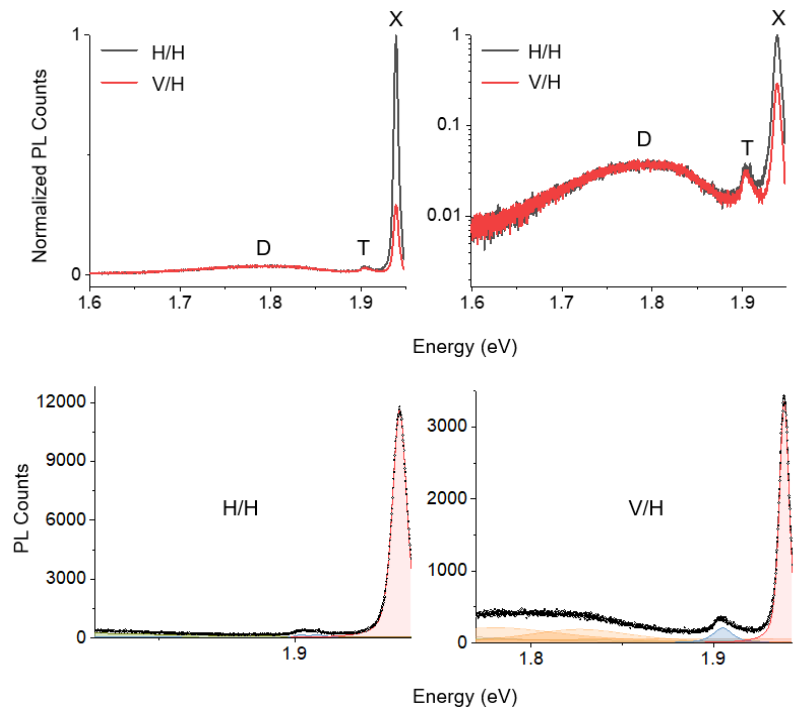

Spectrum in main manuscript Fig. 3f:

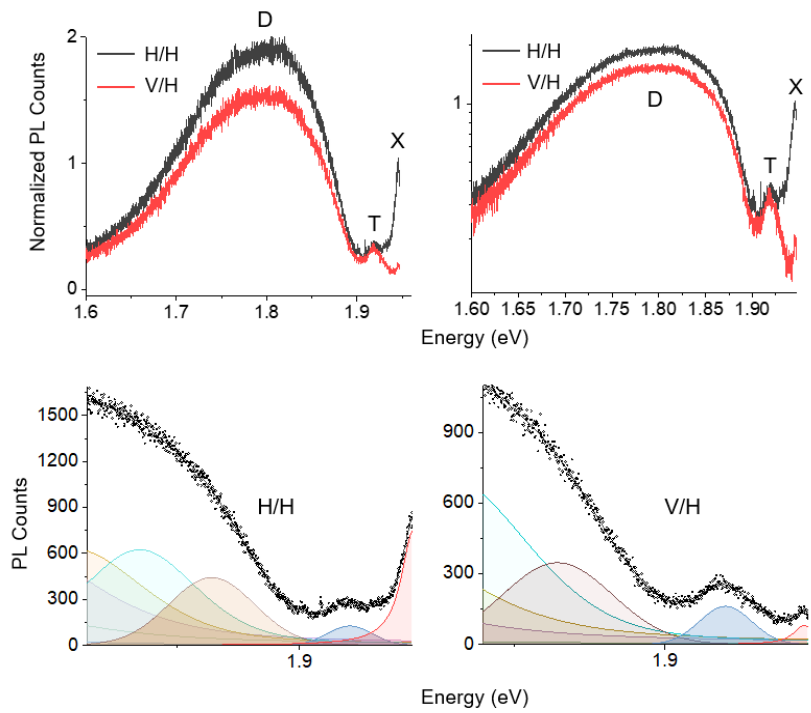

Spectrum in main manuscript Fig. 4b:

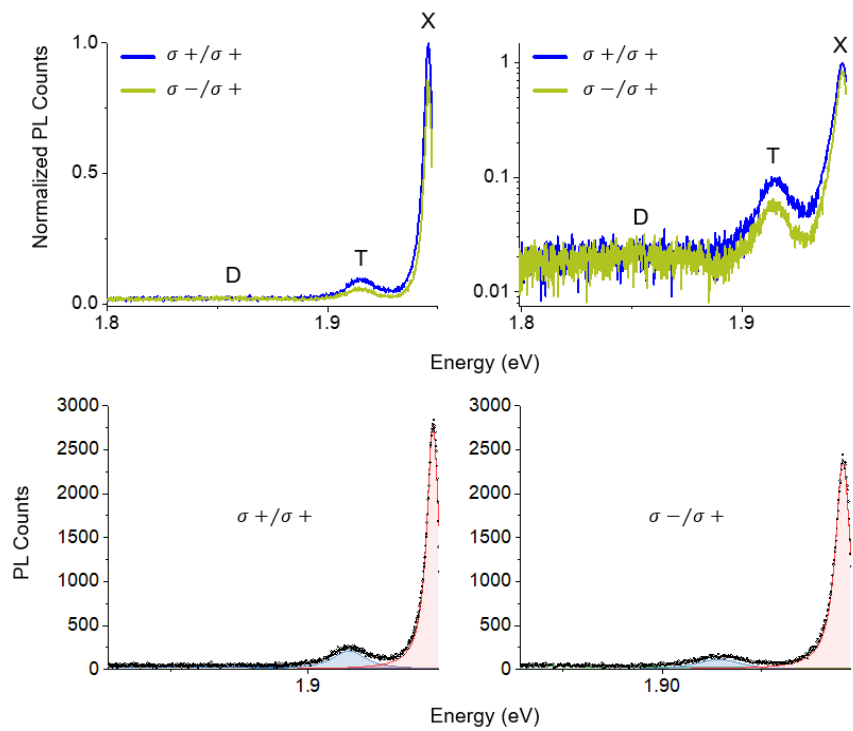

Spectrum in main manuscript Fig. 4c:

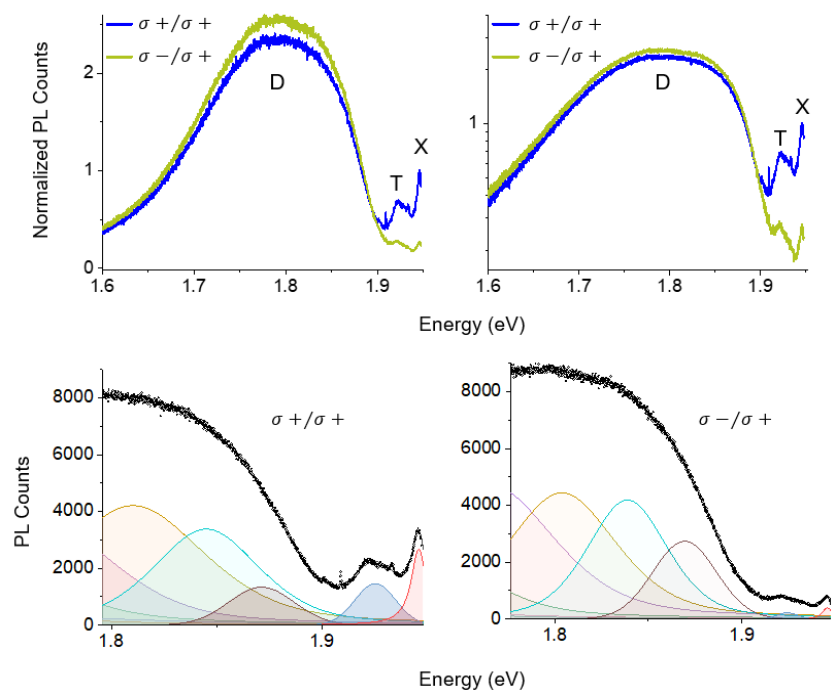

**Note 12.**

**Polarization contrast as a function of momentum scattering rate at varying exciton lifetimes**

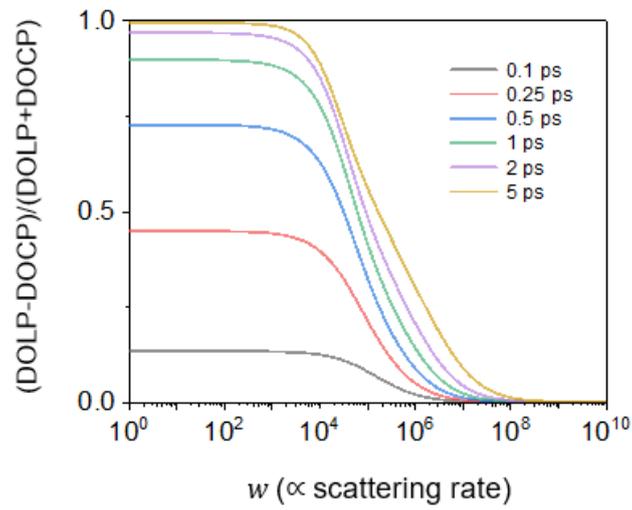

**Fig. S12.1:** Polarization contrast [PC = (DOLP-DOCP)/(DOCP+DOLP)] plotted as a function of momentum scattering rate at different exciton lifetimes.